\documentclass[review, preprint]{elsarticle}

\usepackage{amssymb}
\usepackage{amsmath}
\usepackage{amsfonts}
\usepackage{tabularx}
\usepackage{graphicx}
\usepackage{fontawesome}
\usepackage{xcolor}
\usepackage{soul}
\usepackage{url}

\journal{Journal of Systems and Software}

 \newenvironment{quote1}{%
   \list{}{%
     \leftmargin 0.25cm   
        \rightmargin 0.25cm
   }
   \item\relax
}
{\endlist}

 \newenvironment{quote2}{%
   \list{}{%
     \leftmargin\parindent
        \rightmargin\parindent
   }
   \item\relax
}
{\endlist}

\newcommand{\myquote}[2]{\begin{quote1}\emph{#1}\end{quote1}} 
\newcommand{\rqquote}[1]{\begin{quote2}{#1}\end{quote2}} 

\begin{document}

\begin{frontmatter}



\title{Streaming Software Development: Accountability, Community, and Learning}

\author[inst1,inst2]{Ella Kokinda}

\affiliation[inst1]{organization={Corresponding Author},
    email={ekokind@clemson.edu}}

\author[inst2]{Paige Rodeghero}

\affiliation[inst2]{organization={Clemson University},
            city={Clemson},
            state={South Carolina},
            country={United States of America}
            }

\begin{abstract}
    People \textcolor{black}{use the} Internet to learn new skills, stay connected with friends, and find new communities to engage with. \textcolor{black}{Live streaming platforms like Twitch.tv, YouTube Live, and Facebook Gaming provide a place where} all three of these activities intersect and enable users to live-stream themselves playing a video game or live-coding software and game development, as well as the ability to participate in chat while watching someone else engage in an activity. Through fifteen interviews with software and game development streamers, we investigate why people choose to stream themselves programming and if they perceive themselves improving their programming skills by live streaming. We found that the motivations to stream included accountability, self-education, community, and visibility of the streamers' work, and streamers perceived a positive influence on their ability to write source code. Our findings implicate that alternative learning methods like live streaming programming are a beneficial tool in the age of the virtual classroom. This work also contributes to and extends research efforts surrounding educational live streaming and collaboration in developer communities. 
\end{abstract}


\end{frontmatter}


\section{Introduction}
\textcolor{black}{Live streaming platforms like Twitch, YouTube Live, and Facebook Gaming typically cater to live streaming video games and phenomena like \emph{Just Chatting}, where streamers get in front of the camera to talk and interact with chat members. From 2020 to 2021, Twitch saw a 40\% increase in watch hours with 24 billion watch hours by the end of 2021, Facebook Gaming viewership rose 47\% to 5.4 billion watch hours, while YouTube Live viewership decreased 13.1\% with 6.2 billion watch hours in 2020 to 5 billion in 2021 \cite{chase2021, chase2022, chase2022_1, may2022}. Combined, these live streaming platforms have over 49 billion watch hours from 2021 throughout the first half of 2022 \cite{may2022}. Live streaming is not limited to playing video games, with streams for musicians and other artistic creatives live streaming themselves paint, sew, and craft becoming more common. Finally, there are categories of streamers where one can watch software development and programming, which we focus on in this paper.} With the move to online learning becoming more commonplace and \textcolor{black}{live streaming platforms} making it easier to stream to a broad audience, we aim to understand this small, yet active software development community's sustainability and purpose for developers \cite{hiltz2005education}.

Much of the recent research into \textcolor{black}{live} streamers and viewers relates to playing and watching video games. However, we believe that parallel research can be applied to the relationships between streamers and viewers in a learning context as well. Wohn \emph{et al.} explores how monetary donations affect social relationships between streamers and viewers, finding that viewers donate to streamers to compensate for learning from the streamer about a particular game, as well as Wang \emph{et al.}'s study on the impacts of gift-giving from the audience to streamers \cite{wohn2018explaining, wang2019love}. It is important to note that while our work does not focus on monetary compensation, it is an element that has not been fully explored in \textcolor{black}{live streaming} research outside of gaming and entertainment contexts and implications for social relationships that might differ from traditional learning settings. 

There is only a small body of work surrounding \textcolor{black}{live streams} and software and game development. \textcolor{black}{As viewership numbers and development live streaming communities continue to grow,} it naturally begs the \textcolor{black}{question} of why are developers streaming? In this paper, we aim to expand upon prior work with a broader and more diverse audience due to the expansion of \textcolor{black}{live streaming} user bases in recent times \cite{faas2018watch, faas2019looking, chen2021towards}. \textcolor{black}{These foundational prior works enabled us to set up our study to expand and deep dive into the motivations surrounding software and game development live streaming. Our work aims to increase representation across all streamers through sampling from streams with varying viewership numbers, not just the top-level streamers with most viewers, and sampling from outside of standard viewership times. Through this we can arrive at a more generalizable and inclusive sample that could be used by developers who might be interested in streaming or viewing live streams.} 

Our work is motivated by the desire to understand the benefits of live streaming in software development and provide design implications for the platform, which could improve and help foster the learning communities that form around these types of social learning interactions. Design implications have the potential to impact the platform\textcolor{black}{s} to make them more accessible to those who want to learn and teach, alongside bringing a better sense of community and accessibility to a platform\textcolor{black}{s} perceived as a place solely for \textcolor{black}{live streaming} gaming. 

\textcolor{black}{Educational and software development live} streamers hosted on these platforms offer accessible and free education for those wanting to learn or explore something new without heavy investment. Therefore, our study explores why streamers want to use \textcolor{black}{live streaming} as a learning vector and attempts to measure perceived software development ability improvements. The contributions of this work also provide knowledge and context to a newer medium of collaboration and learning for software and game development.      

In this paper, we focus on understanding the motivations of software developers who stream themselves programming, work to understand if these streamers perceive an improvement in their software development skills, and explore the benefits and limitations of the \textcolor{black}{live streaming} platforms identified by streamers. Exploring the interview data, we found that streamers often have a community following that motivates and holds them accountable to stream and continue developing their work. This idea of a community also influences the idea of self-education and using streaming to explore new ideas or practice known software development skills. Finally, we discuss future considerations about expanding this study and where preliminary data shows potential for additional research. 

This paper makes several contributions to software engineering education and expands knowledge around the benefits of live streaming, online education, and developer communities \textcolor{black}{by:}
\textcolor{black}{
\begin{itemize}
    \item investigating and providing findings on initial and continued motivations to live stream software and game development
    \item discussing the beneficial outcomes of live streaming development while taking into account streamers' initial and continued motivations
\end{itemize}
}

We believe live streaming software development provides developers with the opportunity to learn and practice accountability and find communities that positively encourage engagement and collaborative learning. We highlight that social interaction and accountability are significant factors influencing developers who choose to live stream and how this may inform software and game development education practices and recommendations for new learners and programming veterans alike. Our work also extends to the information-seeking practices of developers and their social behaviors. 

\section{Background \& Related Work}
In this section, we discuss the background and related work on \textcolor{black}{live} educational streams, interactions of streamers and viewers in \textcolor{black}{live streams'} integrated chat feature and within communities that form around streamers, streamer motivations, and limitations within prior work. 

\subsection{Twitch and Educational Streams}
As the Twitch platform evolves with additional categories of streams outside of video games, the categories of \emph{IRL}, \emph{Just Chatting}, and \emph{Educational} have more traction with viewers \cite{steinbeck2021teaching}. A 2018 study focused on video game programming streams indicated that this form of live streaming has the possibility to be integrated into mainstream online learning environments \cite{faas2018watch}. Live streaming has some benefits over prerecorded and structured educational programming videos. Chen \emph{et al.} found that live streaming has benefits over prerecorded videos due to less upfront planning and financial commitments, which gives streamers a lower barrier of entry to teaching \cite{chen2021towards}. The ease of getting onto the streaming platform means that it is accessible to those who may not have formal pedagogical training but can still share information and knowledge with others. Streamers are then not limited to traditional ways of becoming a professor or teacher. Prior \textcolor{black}{research} surrounding community and developers' work has shown that streaming platforms benefit developers as a means of developer advocacy and providing insights into software development careers~\cite{chattopadhyay2021developers}.

For viewers, streaming was found to be beneficial for those who preferred an ``over the shoulder'' style of learning, informational learning, and interest in seeing how others approach a problem set \cite{chen2021towards}. This style of informal learning that Twitch provides has also been shown to give learners the ability to personalize their learning goals and interests \cite{selwyn2007infromal}. However, a detriment to viewers, the Twitch platform does not go any further to categorize streams. Viewers cannot search for specific styles of teaching or, in some cases, specific areas of interest, nor does the platform afford that type of interaction style for content creators \cite{chen2021towards}. The platform attracts those who are seeking information, even in the context of watching others stream video games, with information-seeking positively impacting usage (time spent watching content) \cite{sjoblom2017people, fraser2019sharing}.

A 2017 experiment with educational streams found that learning performance for viewers of educational streams is roughly the same for being taught by a beginner or an expert \cite{payne2017examining}. This study also found that the personality of the streamer factors into learning performance - viewers who found streamers agreeable have a positive takeaway from new streamers \cite{payne2017examining}. These streamer attributes factor into how and why communities of viewers may form. 

\subsection{Streamers and Interactions with Chat}
Hosting a stream on Twitch \textcolor{black}{and other platforms like YouTube and Facebook Gaming} is free; there is no cost to the streamer concerning hosting themselves on the platform, same as many prerecorded video hosting websites like YouTube and Vimeo. No upfront monetary costs make the platform accessible with low barriers to entry. In general, streamers are motivated by a need for companionship and a desire to socialize \cite{hamilton2014streaming}. A 2019 study within low-viewer count programming education streams found that a significant motivation to stream was finding help with their programming and for socialization \cite{faas2019looking}. This same study also found that these streamers were in a transitional state in life that enabled them to stream regularly and gain access to a place to socialize or a community with similar interests~\cite{faas2019looking}.

One of the core elements of the \textcolor{black}{live streaming} platforms is the live chatting feature. During streams, a live chat continuously updates with messages from viewers to the streamer, streamer to the viewers, and amongst the viewers. These three types of interactions facilitate community development. Within well-established communities, mentorship is found to be three-fold, mirroring the types of interactions that the platform affords \cite{faas2018watch}. This indicates that interactions facilitate learning within these communities by the streamer teaching or working through a concept, a viewer pointing out a potential bug, and viewers explaining concepts to other users. Communities around a streamer manifest in a few ways with regulars and off-site communities due to \textcolor{black}{live streaming platforms} not having \textcolor{black}{adequate or apparent} features to facilitate discussion outside of a stream when it is not live. These communities can be found on Patreon\footnote{A crowdfunding platform to pay people for content they produce; https://www.patreon.com/}, Discord, Reddit, and others. 

While community and mutual benefit are positive aspects of streamer and viewer interactions, some downsides exist. Streamers have and continue to voice concerns over harassment, privacy, and moderation tools available to them \cite{chen2021towards}. Volunteer moderators (mods) can monitor chat for inappropriate, hateful comments or spam; however, this responsibility typically falls on the streamer to moderate their chats while live. This downside to the platform dissuades new streamers from choosing to start streaming or is the reason why they discontinue streaming. \textcolor{black}{Many live streaming platforms} have not addressed community concerns for better moderation, better ways for viewers to find the content they seek, nor provide better tools to enable creative streamers to engage their viewers in new, more meaningful ways \cite{fraser2019sharing}. Additionally, prior work lays the foundation for understanding that streamers, though they might know the potential downsides of harassment, will use the platform to overcome social anxiety and use the negative aspects of the platform as exposure therapy \cite{frommel2021potential}.

\subsection{\textcolor{black}{Limitations \& Gaps in Literature}}
Limitations exist in prior studies in accounting for all genders for both streamers and viewers \cite{faas2018watch, faas2019looking}. Very recent research has more diverse demographics \cite{chen2021towards}, but there still is a gap in representation. \textcolor{black}{Based on our observations, less than 15\% of streamers in software and game development are female, our work begins to reach a more representative sample with 20\% of our participants identifying as female or non-binary but female-presenting.}

\textcolor{black}{Additionally, while prior work assumes many of the motivations to live stream are the same as motivations for gaming streams \textcolor{black}{or omits development streams from analysis}, our work differs by directly investigating developers' initial and continued motivations to live stream. This larger scope of why developers choose to stream contextualizes the educational benefits of development live streams, provides new data on the beneficial outcomes of live streaming development for streamers and viewers, and extends and confirms knowledge on the benefits of live streaming developer communities} \textcolor{red}{\cite{faas2018watch, fraser2019sharing}.}

\textcolor{black}{Finally, this paper does not investigate the implications or motivations of financials or monetary gain for streamers, nor the motivations for viewers to give monetary gift donations to streamers. Prior work regarding monetary compensations for non-development streams, often on Twitch through ``bits'', advertisement revenue, or another funding platform (\emph{e.g.,} Ko-fi, Patreon, Gumroad, Paypal), indicates that there may be a link between compensation and motivation for streamers \cite{wohn2018explaining, wang2019love, johnson2019and}. Additionally, while we argue that streaming software and game development has a low up-front cost and low barrier to entry, recent work investigates investing in stream peripherals and equipment aids in audience engagement and interaction \cite{drosos2022design}.}

\section{Methodology}
In this section, we describe our research questions and the rationale behind them, our study design, participants, data collection, and data analysis. 

\subsection{Research Questions}
The goal of this work is to better understand the motivations, intentions, and perceptions \textcolor{black}{of software and game development streamers}, understand streamer and viewer interactions from the streamer's perspective, the perceived learning expectations of streamers, \textcolor{black}{streamers' perceived learning outcomes for viewers}, and design implications for \textcolor{black}{live streaming platforms} that could enable more people to utilize the platform for educational purposes. Therefore, we ask the following Research Questions (RQs):

\rqquote{{$RQ_1$:} What are the motivations of developers that live stream?}
\rqquote{{$RQ_2$:} What are streamers' perceptions of their viewers and how do they perceive their engagement with viewers?}
\rqquote{{$RQ_3$:} What are developers' perceived \textcolor{black}{beneficial outcomes} of live streaming for themselves and for their viewers?}
\rqquote{{$RQ_4$:} What are the challenges developers have while live streaming?}

The rationale behind \emph{$RQ_1$}, \emph{$RQ_2$}, and \emph{$RQ_3$}  is to understand a holistic perspective of the streamer about themselves and their motivations, as well as their perceptions about the viewers of their stream. This information helps us to understand more about programming education and developer communities. From there, \emph{$RQ_4$} addresses challenges that can help influence platform design choices and connects back to the platform as an educational and community-forming media for those interested live streaming their programming.

\subsection{Interview Design}
We designed our study to elicit qualitative data through semi-structured interviews. We interviewed participants for 40-70 minutes about their experiences and thoughts on streaming, live coding practices, perceptions of their viewers, and thoughts on the \textcolor{black}{live streaming} platform. We started each interview by collecting demographic information and moved into a semi-structured and open-ended question-answer format. We asked the participants about their programming and educational experience, stream-specific questions, interactions with viewers, and then ended the interview with questions about their perceptions of their viewers and any community they may have formed through streaming.

\subsection{Methods} 
The authors identified interview participants by browsing live streams on Twitch between the hours of 6 a.m. and 12 a.m. Eastern Standard Time (\textcolor{black}{EST}) for seven months, selecting streamers at random from who were at the top of the \textcolor{black}{\textit{Software and Game Development}} category (most current \textcolor{black}{viewers}) to the bottom of the category streamers with low viewer counts (typically less than five viewers), noting down usernames and using the Twitch platform to identify preferred means of contact. \textcolor{black}{These broad hours were chosen to reach international streamers and those who may not fall into typical popular streaming times\footnote{\textcolor{black}{Popular streaming times vary, but typically fall between 3 a.m. EST and 11 a.m. EST, and again from 2 p.m. EST to 5 p.m. EST. \cite{may2022_1, nielsen2022}}}, and to sample from streams with varied viewership numbers.} For each streamer, we would observe the live stream for 15-30 minutes to ensure that the content presented was \textcolor{black}{primarily English-speaking, and} relevant and on-topic to software or game development. \textcolor{black}{During our observations, we would note down aspects of the stream's presentation and experience (\emph{i.e.,} overlays, interactive elements), engagement of the audience, and streamers' disposition. However, we do not take these elements into account in our analysis as they would fit better into further research into stream communities and nuanced research from viewer perspectives. We excluded streams where the streamer was not present (\emph{i.e.,} an empty chair with no ``Be Right Back" screen), streams with non-software and game development-related content, non-English speaking streams, and overlaid content blocked the primary view of the stream content.} We noted subscriber counts \textcolor{black}{and current viewers} directly from their Twitch page, and used TwitchTracker\footnote{A third-party Twitch streamer statistics tracker. https://twitchtracker.com/} to gather historical viewership averages.

We conducted and recorded interviews between February and September 2021 using Zoom and Discord voice calls, recorded with participant permission, and anonymized transcripts. We present a sample of interview questions in Table \ref{table:interviewQs}. No incentives were offered to participants to complete the interview. Interviews began with basic demographic information before moving into general streaming questions, questions specific to participants' streams, and finally, questions about their interactions with chat.

\subsection{Participants}
 In total, we sent 103 recruitment messages through email and Twitter direct messaging. We collected 15 responses for participation (response rate 14.5\%). Of 15 participants, 2 were females, 12 were males, and one was non-binary. The age of participants ranged from 23 to 55 years old, with a mean age of 34 and median age of 31. Participants resided in the United States, Germany, the United Kingdom, Australia, or Canada. Most participants (11 of 15) indicated they were self-taught programmers. \textcolor{black}{Interestingly,} 9 of \textcolor{black}{these} 11 \textcolor{black}{self-taught participants had indicated prior} formal computer science or STEM-related education. \textcolor{black}{Self-taught participants, like P1, noted they learned \emph{``development fundamentals''} through work experience post graduation, same with P3 who entered the workforce as a developer before beginning his formal STEM education. Participants like P6, P11, P12, and P14, all noted that they used online resources and personal projects to teach themselves software development and other programming skills. Additionally, participants like P2, P5, P10 and P11 expressed interest in hobby programming before engaging in formal education.} Programming experience ranged from 3 to over 25 years of experience; 1 participant had less than 5 years of experience, 4 participants had 5-10 years of experience, 3 had 10-15 years of experience, 2 had 15-20 years of experience, and 5 participants had more than 20 years of experience. \textcolor{black}{Streaming experience ranged from less than a year to 8 years of experience, with five participants noted having streaming experience before starting their software and game development streams.} Table \ref{table:demos} presents a breakdown of participant demographics with participant identifiers (PID) and \textcolor{black}{stream} statistics. Please note, we do not provide all demographics or \textcolor{black}{stream} statistics to protect participants' anonymity.

\begin{table*}[t!]
\centering
\resizebox{\textwidth}{!}{\begin{tabular}{ccccccccccc}
\hline
PID &
  Age &
  Gender &
  Location &
  Education &
  \begin{tabular}[c]{@{}c@{}}Self-taught vs.\\ Formal Education\end{tabular} &
  \begin{tabular}[c]{@{}c@{}}Programming\\ Exp. (Years)\end{tabular} &
  \begin{tabular}[c]{@{}c@{}}\textcolor{black}{Streaming} \\ \textcolor{black}{Exp.} (Years)\end{tabular} & 
  \begin{tabular}[c]{@{}c@{}}Full vs. Part-time\\ Streaming\end{tabular} & 
  Subscribers & 
  Avg. Viewers \\ \hline
P1  & 33  & male       & US             & BS in CS, PhD \textcolor{black}{in Econ}          & Self taught                                                                & 18                                                                 & 8\textcolor{black}{*}                                                                                     & Part-time                                                                                  & 6,900                           & 277             \\
P2  & 23  & male       & US             & BS in CS     & Self taught                                                                & 8                                                                  & 7\textcolor{black}{*}                                                                                        & Part-time                                                                                  & 1,400                           & 10              \\
P3  & 29  & male       & US             & BS in CS     & Self taught                                                                & 19                                                                 & 3                                                                                     & Part-time                                                                                  & 15,700                          & 68              \\
P4  & 44  & male       & Germany        & PhD          & Not described                                                              & 20                                                                 & 1.5                                                                                   & Part-time                                                                                  & 1,200                           & 103             \\
P5  & 42  & male       & US             & BS in CS     & Self taught                                                                & 20+                                                                & 2                                                                                     & Part-time                                                                                  & 3,500                           & 20              \\
P6  & 39  & female     & United Kingdom & High School  & Self taught                                                                & 9                                                                  & 2                                                                                     & Part-time                                                                                  & 4,500                           & 46              \\
P7  & 29  & male       & Australia      & BS in HCI    & Not described                                                              & 15                                                                 & 4                                                                                     & Full-time                                                                                  & 26,000                          & 79              \\
P8  & 31  & female     & US             & PhD \textcolor{black}{ in CS}         & Not described                                                              & 13                                                                 & 1.5                                                                                   & Part-time                                                                                  & 503                             & 9               \\
P9  & 30  & non-binary & US             & Art Degree   & Not described                                                              & 9                                                                  & 4\textcolor{black}{*}                                                                                        & Part-time                                                                                  & 4,100                           & 17              \\
P10 & 24  & male       & England        & BS in CS     & Self taught                                                                & 15                                                                 & 1.5                                                                                   & Part-time                                                                                  & 256                             & 6               \\
P11 & 55  & male       & US             & High School  & Self taught                                                                & 20+                                                                & 3                                                                                     & Part-time                                                                                  & 1,400                           & 7               \\
P12 & 43  & male       & US             & Some college & Self taught                                                                & 25+                                                                & 2                                                                                     & Part-time                                                                                  & 2,300                           & 17              \\
P13 & 23  & male       & US             & Some college & Self taught                                                                & 3                                                                  & 4\textcolor{black}{*}                                                                                        & Part-time                                                                                  & 7,400                           & 42              \\
P14 & 39  & male       & US             & BS in CS     & \textcolor{black}{S}elf taught                                                                & 25+                                                                & 5\textcolor{black}{*}                                                                                        & Full-time                                                                                  & 8,600                           & 24              \\
P15 & 26  & male       & Canada         & BS in CS     & \textcolor{black}{S}elf taught                                                                & 7                                                                  & 0.75                                                                                  & Full-time                                                                                  & 205                             & 7               \\ \hline
\end{tabular}}
\caption{Streamer Interview Participant Demographics \\\textcolor{black}{* - indicates participant had prior experience streaming before development live streams}}
\label{table:demos}
\end{table*}

\subsection{Data Collection} 
The data that we collected includes the streamers' programming background with years of experience and languages they commonly use on and off stream, educational background, work experience as a developer, and time they have spent streaming. Additional data we collected includes insights into streamers' motivations to start and continue streaming, preparation for a stream (\emph{i.e.,} if preparing for a stream, how?), and notable \textcolor{black}{successes} and challenges of streaming as a whole. To understand the interactions a streamer has with chat\textcolor{black}{,} we collected data regarding the frequency of interaction with viewers, what a typical interaction with a viewer entails, how the streamer interprets chat suggestions for programming errors or bugs, and if they have any routine regulars to their streams they recognize. Taken together, streamers' development backgrounds, streaming motivations, and interactions with chat members and viewers provide a holistic high-level understanding of how and why developers use live streaming.

\begin{table*}[]
\centering
\resizebox{\textwidth}{!}{\begin{tabular}{l}
\hline
\textcolor{black}{Background Questions (B) and} Subset of Interview Questions \textcolor{black}{(Q)}\\ 
\hline
\textcolor{black}{\textbf{B1:} Can you describe your background education? Do you have a formal computer science or STEM-related background? Do} \\ \textcolor{black}{you have any formal education in teaching?} \\
\textcolor{black}{\textbf{B2:} How long have you been programming?} \\
\textcolor{black}{\textbf{B3:} What is your current work background - are you employed as a software developer or as an educator?} \\
\textcolor{black}{\textbf{B4:} What languages do you typically code in?} \\
\textbf{Q1:} What made you start streaming? How long have you been streaming for?\\
\textbf{Q2:} What motivates you to continue to stream?  \\
\textbf{Q3:} Do you watch other educational or Science and Technology Streamers\textcolor{black}{?} If so, do you pull any influence \\ or inspiration from them? \\
\textbf{Q4:} When streaming software development, how do you go about describing what you are doing to viewers?  \\ 
\textbf{Q5:} Are there any successes as a streamer that stand out to you? Any challenges or demotivators? \\
\textbf{Q6:}  How much chat interaction is there in general on your streams? Can you speak to any examples of chat interactions? \\
\textbf{Q7:}  Does chat ever make suggestions for improvement or point out bugs in your code? How do you handle this? How do you \\ feel about them doing that? \\
\textbf{Q8:} Do you have any stream regulars? Are you able to gauge their level of programming experience based on interactions\\ you may have with them? \\
\textbf{Q9:} Do you have a community set up outside of your streaming platform for your followers? \\
\hline \\
\end{tabular}}
\caption{\textcolor{black}{Background Questions (B) and} Subset of Streamer Interview Questions (Q)}
\label{table:interviewQs}
\end{table*}

\subsection{Analysis}
We treat interview transcripts as qualitative data and analyzed them using thematic analysis and qualitative analysis guidelines outlined in McDonald \emph{et al.} \cite{mcdonald2019reliability}. Using the main themes from the initial three interviews, we identified similar and additional themes in the following data and organized them into broader themes to identify any relationships and reoccurring themes between the participants and of live streaming development as a whole. During the data collection, the first author would revisit earlier interviews to assess and identify any new potential themes missed that were realized during the analysis of subsequent data and correlate consistent themes between the interviews. 

Upon completion of the interviews, we reviewed the data and, with high-level themes, identified and organized sub-themes related to motivations to stream, audience members' perceptions, skill improvement, and challenges of streaming. The authors met to iterate, discuss, and refine themes and sub-themes. 

\section{Results}

In this section, we discuss the results of the developer streamer interviews.  We outline our identified themes within the primary motivations of streaming, perceptions of audience members, perceived self-education and skill improvement, and notable comments about the challenges of streaming and platform difficulties from software and game developers \textcolor{black}{on streaming platforms}. 

\subsection{\textbf{$RQ_1$:} Developer Motivations to Stream}
To understand developer motivations to stream ($RQ_1$), we asked streamers questions focused on why they chose to stream, how they became involved in streaming, and their continued motivations to live stream. Though programming experience, education, and years of streaming varied among the participants, they shared much of the same starting motivations, continue to share some of the same starting motivations, and have evolved as their streams have matured. We found that starting and continued motivations converge around five themes - (1) accountability, (2) self-education, (3) teaching and mentoring others, (4) visibility of the streamer's work, and (5) finding a community with shared interests. \textcolor{black}{Throughout} this section, we discuss the themes and present quotes from our participants. Table \ref{tab:themes} presents a high-level overview of each theme we identified for $RQ_1$ with top-level attributes of what the theme entails. 

\begin{table*}[]
\centering
\resizebox{\textwidth}{!}{\begin{tabular}{ll}
\hline
Theme                                                                              & Definition                                                                                                                                                                                                                                    \\ \hline
\textit{Accountability}                                                             & Setting aside time to work, making a keeping a schedule or routine, fulfilling obligations to others                                                                                                                                          \\
\textit{Self-education}                                                            & learning a new programming language, keeping current skills through practice or challenging problems                                                                                                                                          \\
\textit{Teaching/Mentoring}                                                        & \begin{tabular}[c]{@{}l@{}}developer advocates for specific languages, sharing programming knowledge as a senior- or high-level developer\end{tabular}                                                                                     \\
\textit{Visibility}                                                                & \begin{tabular}[c]{@{}l@{}}advertisements for their games, insights into development processes, directing off \textcolor{black}{streaming platforms} to ways for \\ viewers to be involved or contribute to a streamers work\end{tabular}                                     \\
\textit{\begin{tabular}[c]{@{}l@{}}Community and \\ shared interests\end{tabular}} & \begin{tabular}[c]{@{}l@{}}finding like-minded individuals interested in a specific programming language or within game development, \\ finding a space where communication and ideas are shared openly with positive intentions\end{tabular} \\ \hline
\end{tabular}}
\caption{Overview of identified themes presented for $RQ_1$}
\label{tab:themes}
\end{table*}

\vspace{0.1cm}\noindent\textbf{Theme 1: Accountability --}
We found that eight of the 15 participants indicated accountability as a beginning or continued motivation to stream (P2, P3, P5, P6, P8, P10, P12, and P14), and was the most common motivation to stream. Streaming provided an opportunity to set expectations for themselves and helped set aside time to work. This also factors into self-education motivations for these streamers and other participants we interviewed.

\myquote{``I was working on a side project. And basically, before doing that I was just sitting around not doing much. So streaming is essentially just a way to hold myself accountable to doing the work and makes it a bit more sociable.'' \emph{-P10}}{}

\myquote{``Live programming on Twitch was a great form of accountability for learning new stuff. And so like, my initial motivation for doing this was basically, I knew I wanted to kind of more regularly do some practicing. And, you know, I didn't really care how many people showed up, but this is going to be my, my public commitment to do actually working on some of this stuff.'' \emph{-P8}}{}

We found streamers held personal accountability and responsibility to provide to the community they fostered \textcolor{black}{while streaming}. Streamers reported that the community they built motivates them to hold and keep a schedule out of respect for their followers, saying: 

\myquote{``I feel a little bit of a responsibility as well. I've got kind of, yeah, like a community of people who, you know, they expect on, you know, during my stream schedule, hours, they expect me to be there. And, and I expect to be there. So, it's a bit of a contract at this point.'' \emph{-P7}}{}

\myquote{``So if you build a community and just keep a consistent schedule, that's super helpful. And both, you know, boosting morale for everybody. And, you know, accountability.'' \emph{-P14}}{}

\vspace{0.1cm}\noindent\textbf{Theme 2: Self-education --}
We found that twelve of the fifteen participants indicated self-education or a desire to learn and practice new concepts by streaming (P2, P3, and P5-P14). Six participants indicated that self-education or learning a new development concept or language was a starting motivation. P12 expressed wanting to learn a new language with accountability and education as motivators. He treated it like learning a new spoken language, where one needs to commit to practice every day in order to get good. Others, who also wanted to teach and mentor others, found a symbiosis in streaming to explore a new concept and have the opportunity to learn from viewers who are active in the chat when they are stuck with a problem on stream. P11 alluded to this back-and-forth learning with: 

\myquote{``Sometimes if I get stumped, rather than Google, people will just chime into chat and say try this. And that interaction that back and forth is, is really, you know, it's almost like being in a room of people. And they're just, you're shouting out, I'll try this, click here, type this. And, and I like that. I like that. It makes it much more like a pair programming type of thing.'' \emph{-P11}}{}

P6, P9, and P10 also shared experiences where they learned from their audience who used the text chat to explain a concept, offer a possible solution, or point out a bug in the code.

\vspace{0.1cm}\noindent\textbf{Theme 3: Educating and Mentoring Others --}
Nine of the 15 participants indicated that one of their motivations in streaming was teaching or motivating and mentoring others through their own work. Within this group, we have two subgroups of streamers: 
\rqquote{{Streamer Group 1} - Those who actively seek to teach and educate others about a development topic.}
\rqquote{{Streamer Group 2} - Those who want to share their knowledge and be a resource for others while not necessarily setting out with the expectation to teach.}

Additionally, we found a subset of streamers in both groups who wanted to be a mentor and advocate for developers. One participant, P4, made teaching and mentoring others his main motivation for streaming, saying they wanted to offer ``free teaching.'' This desire came as many schools moved online due to the pandemic and used it to give back to people who might be missing out on in-person opportunities or were taking the time at home to learn something new. Having a background in education made the transition easy, and they translated their work into a multi-day workshop where people could come and learn as though it was a real class setting. They (P4) also streamed weekly for different development concepts while offering the multi-day workshop/class once or twice a year to make what they do \emph{``open and freely accessible, and not locked away, behind paywalls''} as they felt other online educational content did.

Other participants, with secondary motivations for teaching and mentoring others, wanted to share the knowledge they have because they think they are doing something unique, have years of experience as a developer and want to share with the next generation of developers, or they want to teach others while teaching themselves something new. P11, an experienced software developer with more than 20 years of experience, explained that he believed many development tutorials leave out the challenges and struggles a developer faces when trying to accomplish something and explained his process of streaming as:

\myquote{``If it took me a long time to do something, why not share that? So other people don't have to spend a long time to do it. That's one of the things that I think I do well on my stream. ...  Here are all the problems I have, here are all the potholes that I almost fell into, versus being something polished and clean. [By] sharing step one through step four, with the potholes and what you ran into between each step, I think it helps a lot of folks.'' \emph{-P11}}{}

Much like P11, P3, with many years of development experience, explained they are passionate about teaching others and will take questions asked on stream, address them live, and then turn them into short YouTube videos for small concepts and items that could help others who potentially missed a stream. P5 also reported that educating others is a motivation because it is a feel-good moment for him when  \emph{``you know, you did a good deed, you're helping out the next version of people''} and seeing them go on and have successes that they bring back and attribute to the streamer. 

To teach and mentor, we found that 14 participants adopted think-aloud techniques to give viewers a stream of conscious thought process and walk them through what the streamer is typically thinking at a high level, unless deliberately prompted by a chat member to explain what is going on with more detail. P7 explained his thought process as \emph{``being able to frame a bunch of information in a way where someone at the end of it knows how to do something''} because he felt it is both beneficial for his development processes and others.

\vspace{0.1cm}\noindent\textbf{Theme 4: Visibility --}
We found that five of the 15 participants indicated that one of their main motivations to stream was the visibility of their work (P1, P7, P8, P14, P15). This mainly applied to game developers using \textcolor{black}{live streaming as a way} to advertise their game in development and other published game-related work. P1 noted that \textcolor{black}{streaming} was the easiest way to get feature releases shown to his community, which is a different take on traditional software applications that typically only give static patch notes. For P15, visibility of his game development was the sole motivator for \textcolor{black}{streaming}, indicating that \emph{``if I didn't have [to] market the game, I wouldn't be streaming.''} For other game developers, their game's visibility was only part of their motivation to stream. Those whose motivations go beyond just visibility, often game developers, stated they used this visibility to drive viewers to other platforms for their content like Patreon, Steam, or YouTube. One participant (P7) noted that without the community they had created and maintained, they would not be able to have visibility on their work and lose out on potential revenue from their work: 

\myquote{``Twitch drives a lot of my content. And I post that content on YouTube. And in other places, and the community, to me that the, my finances and my revenue are based around my community, and that community lives on a bunch of different platforms, but I spend most of my time on Twitch.'' \emph{-P7}}{}

For the sole software developer, P8, who indicated visibility as a secondary motivator, they advocated for a niche programming language and stream to help others understand the language and its role in certain types of development. They also streamed for visibility and advocated inclusively for LGBTQ+ and female representation in software development.

\vspace{0.1cm}\noindent\textbf{Theme 5: Communities and Shared Interests --}
All participants  indicated that having a community and body of people to speak to and who have a shared interest in development was one of their main motivations to stream. Within the realm of communities, participants starting motivation to stream was to find a community and others like them, while other participants did not seek a community but welcomed and were motivated by the community that has grown around them. Community benefits were multi-faceted in fulfilling social interaction, receiving feedback on projects, accountability to provide content for the community, and learning from others to grow development skills.

We found for five participants that being at home drove them to seek out social interaction, community building, and engaging with other developers online through \textcolor{black}{live streaming}. P5, who resides in Redmond, WA, home to many technology companies and influential software developers, wanted connections with developers outside of his area, feeling like they might bring new ideas to his chat and wanted to connect with \emph{``a new audience.''} 

While some sought to create a community for others, one formed around them and became a motivator to stream once established. P3, who reported an established reputation in the Python community-at-large, notes \emph{``I'm going to be doing this work anyway. So I might as well do it while I'm interacting with people,''} indicating that he would be working on development anyways and would rather have an outlet to create and interact with others while doing it, and later stated that ``\textit{the community aspect is a big reason that I do it.}'' Still for others, like P4, whose primary motivation was to teach others by streaming, felt \emph{``like I'm giving something back to the community. And people seem to appreciate this. And that keeps my motivation up. And so it becomes like a kind of reciprocal thing.''} P7 echos many of the same feelings and sees the community he has built not just as a community but as \emph{``friends. And not only do I want to see them and interact with them, and keep them updated on what I'm working on.''} He also reiterated the feeling of reciprocal giving and the ``contract'' he has made with his friends and audience. P9 also emphasized seeing their audience as friends and that streaming is the only way they get to \emph{``see their friends.''}

Community feedback is an important part of streaming. Whether it is pointing out bugs in code, feedback on game mechanics or art style, and providing suggestions, streamers welcomed the interaction and the opportunity to \emph{``rubber duck\footnote{A \textcolor{black}{method} of debugging \textcolor{black}{and problem-solving technique} where the developer thinks aloud to an inanimate or \textcolor{black}{or abstracted} object and explains line by line what they are doing with code; http://lists.ethernal.org/oldarchives/cantlug-0211/msg00174.html}''} with the community. This feedback often leads to self-education and is expanded upon in \emph{Section 5.3 Improving Development Skills} \cite{hunt1999}.

\vspace{0.1cm}\noindent\textbf{\emph{Pivotal Moments of Change in Motivation --}}
While many participants found that their motivations grew and changed slowly over time, others indicated they had a moment where they realized what their motivations had become or what their motivation was. This was overwhelmingly related to having a community and seeing the impacts of their efforts within the community both on and outside of \textcolor{black}{streaming platforms}. Participants noted instances of people coming to them asking for help or critique on their own projects, seeing viewers excel at a topic the streamer helped them with, sharing personal successes with the streamer, being recognized for their online work at a conference, and seeing the community grow and be interested in the work the streamer produces. 

\subsection{\textbf{$RQ_2$:} Perceptions of Viewers}
To better understand the audience of development streams, we asked streamers about their perceptions of their viewers and what they notice and generalize about them. Additionally, we asked streamers their perceptions of stream engagement activities from viewers like backseat coding\textcolor{black}{\footnote{\textcolor{black}{Backseating is a phenomenon where the person or people not in control tell those who are what to do and how to do it, with and without prompting. Backseat coding and backseat coders are those who tell someone who is programming what or how to do something within their code.}}} and contributions to open-source software (OSS) or the streamer's own projects.

Many participants commented on perceiving most of their viewers and followers to be younger than them, from ages 13-18, and that they were relatively inexperienced coders. Others reported they had a mix of age ranges and experience levels, even noting instances when chat members with more experience chime in with bug fixes or alternative solutions to a problem. Participants who believed they had a younger audience shared that they had instances of a viewer thanking them for teaching them about aspects of software development, being a mentor, and inspiring them to pursue programming. \textcolor{black}{P5, noted that he gravitates toward mentoring younger people on his streams who ask novice questions and inquire about what professional development entails:  }

\myquote{``Most of these high school kids, they're looking for something, you know ... helpful, fulfilling [to] them. And they have all the time but they don't really have necessarily, somebody to spend it with, and sometimes they just would rather spend it with me, which is okay. \textcolor{black}{... And so if I can present them a little bit of my world, of some of the things that at least I know fairly well enough, that maybe I can inspire them to do the same sort of things.}'' \emph{-P5}}{}

Another participant (P7), who also noticed younger viewers wanted to have a one-on-one with him, felt these interactions might disrupt his regular streaming habits but would engage carefully to \emph{``never discourage anyone from contributing''} to the stream. We found P11 had a similar experience with a viewer who was \emph{``brand new to programming''} and participated in many of the streams and is now a full-time developer with the help of P11's streams, in addition to other self-teaching resources. This viewer kept in contact with P11 and actively participates in streams. While P11 does not believe that this viewer learned everything from him, he believes he enabled this individual to explore software development and gave him the tools he needed to succeed. We found that the developer advocate for a niche computing interest (P8), spoke about a regular viewer who had only a \emph{``general interest''} in the field and began participating in her streams. This viewer ended up working for a large technology firm in this niche interest a few years later and felt like she gave this person \emph{``the context and general understanding of the field''} over specifics of a programming language. 

While not everyone could definitively share their experiences of viewers' successes, all participants shared instances of viewers' backseat coding and perspectives of how viewers engage with the stream. Many streamers who started small would actively engage with each viewer in chat if they spoke up or asked a question. \textcolor{black}{Engagement from viewers materialized as backseat coding, where viewers would ask why the streamer is implementing something in a particular way, offering solutions as to how they would approach a problem, and pointing out potential problems in the code. Streamers noted these types of interactions were helpful and productive more often than not. Backseat coding offered streamers an easy way to interact with their viewers and would facilitate additional discussion amongst viewers.} 

\myquote{\textcolor{black}{``I had a gentleman that contributes to the [Arch Linux] kernel. He was in my stream, way more knowledgeable about the topic matter than I was. And we basically just became best friends. We're in it to learn together.'' \emph{-P5}}}{}

\myquote{\textcolor{black}{``There was somebody that I know from the community and she chimed in, and she said, why are you doing it that way? And then it ended up being that culminated into about three weeks of us doing a back and forth development on this.'' \emph{-P11}}}{}

\myquote{\textcolor{black}{``I do make mistakes, because I'm human. And [viewer-name] is pretty good about pointing out mistakes in that case. So like, when I'm working on something that's easy enough for people to follow, they'll be, you know, pointing out typos, syntax errors, that sort of thing.''\emph{-P3}}}{}

\myquote{\textcolor{black}{``You can do things 20 different ways. And I do it one way, and somebody suggested, you could also do it another way. And that, to me, were always good suggestions, but it was never that somebody is really coming into critique the way I'm doing things.'' \emph{-P4}}}{}

Participants indicated that engagement with viewers, \textcolor{black}{through backseat coding or prompting discussion} is the best means of growing and continuing to grow their community. For streams with less engagement, P11 would actively call out \textcolor{black}{viewers using chat features} if they saw people were watching but disengaged: 

\myquote{``I constantly call out especially if it's quiet. I always go to the chat window and I click on the drop down and to see users in chat, because you can see users in chat. And and so I bring that up, and I'm like, are there really people there because they're being really quiet. And I'll see that there are people there. So then I'll just go and say, suppose say something quite typical like `Hey, thanks for coming. I'm glad to see there are some of you out there, please say hi.' You know, `let me know if you have any questions.' And sometimes people will chime in.'' \emph{-P11}}{}

Some participants whose streams have gradually changed noted the different types of viewers and how they engage with the stream - those who are there for the technical aspects and those there to consume a general type of content. We found that streamers must balance their interactions with each type of viewer in different ways not to alienate or bore other audience members while still trying to keep both engaged and entertained by their content. P12 shared his experience with changing the content of his stream:

\myquote{``I saw the audience changed when I switched away from my MMO [gaming streams]. So some people were there from that, some people were there for the game development. When I started to do the WordPress alternative [streams], the audience changed. So just deduction says some people were there for the game development, some people were there for other stuff. And you can tell that other people have just been there for all of it, like somebody has been subbing for, like, nine months now. And, like, they're just buying me a coffee, right?'' \emph{-P12}}{}

P7 discussed how even though their content changes because of the aspects of game development, they were remembering usernames in chat and having respect for the viewers who are retained over extended periods of time. 

\myquote{``I'll still remember them when they come in. And those people who are contributing in that sense, either whether it's like, just supporting me supporting the project from wanting to be here kind of thing or inputting ideas, those people I kind of always have time for, and I always have like a pocket of my brain that I can like, yeah, I can always address them and they respect the amount of time they take up as well.'' \emph{-P7}}{}

Viewer retention is an essential aspect for many participants as it helps gauge the growth and health of their community. Additionally, retention of viewers provides positive feedback about their stream if people return consistently over time. For those that return, streamers echoed the sentiment of P7 that recurring viewers have an investment in what the streamer is providing. 

\subsection{\textbf{$RQ_3$:} Live Streaming Benefits - Skill Improvement}
We asked participants questions about their successes encountered while streaming and how those events affected them personally and within their streams. As many used streaming to hold themselves accountable or learn new concepts, the main benefits we observed surrounded improved programming or development abilities. Additionally, to understand this benefit for streamers, we asked participants if they perceived viewers experiencing programming skill improvements by watching their live streams.

\vspace{0.1cm}\noindent\textbf{\emph{Streamer Skill Improvements --}}
As discussed in Section 5.1, most participants sought to learn or reinforce their development skills through live streaming. We found that streamers expressed challenges while streaming but continued to learn and overcome those development challenges. For example, one participant (P8) encouraged her audience to tell her what they wanted to see and learn about and dedicated several streams to viewer-requested content. 

Another participant (P14), did something similar with a built-in development day for \emph{``Feedback Friday''} where he took feedback from game testers and tried out changes or implemented features that audience members and game beta testers requested. Another participant (P12), after several years of streaming, felt that they would finally be employable in the language they chose to learn on stream. Others who used accountability as their start motivation have continued to stream regularly over several years. 

We found several participants who did not expect to learn from chat members but ultimately found themselves in the situation where they had. P7 noted, and P10 echoed the same sentiment:

\myquote{``I don't have any expectation that I'll learn from the community, but it has proven to be the case where I have done. I mean, I've definitely improved as a programmer, like, a lot.'' \emph{-P7}}{}

\myquote{``People in chat have been very knowledgeable about the thing that I'm doing and have been able to help me.'' \emph{-P10}}{}

We found P11 referring to his interactions and learning from the chat as \emph{``like pair-programming\textcolor{black}{\footnote{\textcolor{black}{A development method where programmers work together at one computer, often where one developer writes the code and the other observes and reviews the code for mistakes, bugs, or gives suggestions. https://en.wikipedia.org/wiki/Pair\_programming}}}''}, especially when faced with a challenge that viewers have potential solutions for or experiences with the same issue. When questioned about his feelings of viewer learning, P13 expresses that it was \emph{``more so [himself]''} learning from chat members and that the roles were reversed. 

\vspace{0.1cm}\noindent\textbf{\emph{Viewer Skill Improvements --}}
We found all of our participants did not have a way to quantify viewer skill development or improvement. However, they shared examples of retained viewers contributing more to live streams, participating in a streamer's community outside of the live streams, and sharing with the streamer successes they have with their projects. 

For some, like P3 and P13, viewers shared their thoughts with streamers like \emph{``your content has helped me become a better software engineer, I have learned techniques that I would not have been able to learn elsewhere just from watching you code''} or simple expressions in chat like \emph{``I understand that now.''} Expressing to the streamers that viewers think they have become better programmers through watching their streams. P6 shared that viewers expressed ``positive feedback'' with her when they were able to learn something from her stream and apply it in their own work. P7 had a viewer come to him for game critiques and feedback on their own game they created. 

Additionally, P2 shared that one of his viewers and long-time community members began to help others in their Discord's designated help channel, as well as contributing to Twitch bots on the streamer's channel. While not distinctly measurable, most streamers felt they positively impacted others, whether that was helping learn a new approach to a problem or providing motivation and \emph{``drive to keep going on their [viewer's] projects.''}

\subsection{\textbf{$RQ_4$:} Challenges of Streaming and Platform Discoverability}
To understand the challenges streamers faced, we asked streamers directly about the challenges they have faced with streaming and while streaming. We found challenges faced by streamers fell into interpersonal or platform-related challenges. To gain deeper insights, we asked questions about what led up to interpersonal friction and what live streaming platforms could do to improve or lessen the impact of challenges.

\vspace{0.1cm}\noindent\textbf{\emph{Interpersonal Challenges --}}
Interpersonal challenges were limited for the majority of the participants. When asked about moderation of their chat, most streamers indicated that it was self-regulating or moderation actions to ban spam bots. We found that all of our female or non-binary participants indicated they occasionally faced interpersonal issues, and one instance was from a male participant. P8, when tagging her streams LGBTQ+, often found her streams attracted \emph{``trolls\footnote{People whose comments are antagonistic, irrelevant, or offensive.}''} and other undesirable comments in the chat. Both P8 and P6 noted that people in chat insulted and harassed them about their intelligence and skills in a particular language. For P8, this was a significant hurdle they needed to overcome and caused anxiety. P6 indicated that other people would try and make streaming competitive and had no great way to deal with it, and it put stress on her and her streams. P11 noted once that a viewer reached out to insult them outside of the \textcolor{black}{streaming} platform after watching a stream where they became stuck on a problem and insulted the streamer about not streaming if they \emph{``don't know it all''} but that this has been an infrequent occurrence. 

While not a result of harassment, some streamers adjusted their stream as they grew. These adjustments were not an easy transition when gaining new viewers and followers. P7 noted that one of the biggest challenges in his streaming career has been adjusting to increased audience size and having more \emph{``backseat coders.''} Interactions scaled from few to many as viewership increased and the drive to interact with all of them was no longer feasible, saying:

\myquote{``The impulse I have is to is to engage directly with all of them, because that's how I've been doing it this whole time. And just from a productivity standpoint, it can just go right off the rails and it actually makes it harder to watch as well. People don't enjoy the streamer getting like sidelined over and over again, trying to address individual concerns or repeating concerns because like, people just jump in, they've just started watching and they haven't got the context of the last 20 minutes.'' \emph{-P7}}{}

Additionally, we found that P3 echoed the same experience of his stream and that his development-to-interaction ratio weighted more toward interactions as his stream grew. He indicated as his community and stream grew, he now \emph{``spends probably 20 to 30\% of my stream actually writing code, and then the remainder of the stream addressing questions from chat or interacting.''} 

\vspace{0.1cm}\noindent\textbf{\emph{Platform Discoverability --}}
Twitch offered one viewership category of \textcolor{black}{Science and Technology} (\emph{S\&T}) when we began this study. Nearly all of the participants had issues with the one category that would define \emph{``science''} as a whole. They often found themselves mixed in with \emph{``duck streams''} of live cameras in duck enclosures and other live services like earthquake monitors. To get around the clutter of S\&T, participants would tag their teams in different ways to try and reach new viewers through tags. However, tags are managed descriptors set by Twitch that streamers can use to describe their streams and that viewers can use to search for streams, which streamers felt limited them in the types of audiences they could reach due to the limited number of tags. Streamers felt that tag searching was not an apparent feature of the website either and that viewers would not use tags to try and find a stream to watch. 

Since beginning our study, Twitch revised categories by separating \textcolor{black}{Software and Game Development} (\emph{S\&GD}) out of \emph{S\&T} into their own category with the same moniker. Many migrated to the new category but expressed hesitation because they built their platform in a different category, and viewers may overlook a new category to search in, as P11, a software developer, noted:

\myquote{``I'm glad they finally put a programming category. But by the same token, nobody's switching to it because it would hurt them. It would make them less discoverable on a platform that's already terrible for discovery.''~\emph{-P11}}{}

Still, with the new category, some game developers expressed that they would like game development to be separate from software development. They felt that while the new category was closer to what they needed, however still felt unseen and less likely to be discovered because game development synthesizes programming and art, as P7, a game developer, explained: 

\myquote{``A lot of the time, I'm working on something that doesn't actually fit into science and technology, it might be concept art, or pixel art or something like that. That becomes more creative. That category doesn't exist anymore. Now, it's like we fit into IRL, or under Science and Technology. There's no way to actually browse for Game Dev, outside of searching for the tag.'' \emph{-P7}}{}

Streaming things like concept art and character modeling felt out of place to game developers in the \emph{S\&GD} category, but felt if they stream in the \emph{Art} category, the replacement for \emph{Creative} category, they would not be as discoverable or lose current audience members when they switched back to programming aspects of the games. The mismatch between appropriate categories and lack of tags for discoverability left game developers feeling it was impossible for them to grow and maintain an audience. \textcolor{black}{Additionally, other platforms like YouTube Live and Facebook Gaming have similarly difficult ways of finding live channels related to software and game development, with tag searching generating prerecorded or irrelevant content.}

\section{Discussion}
Through our interviews, we answered four research questions showing development streams and streamers 1) have multifaceted motivations and benefits for streamers that stem from personal and career-driven goals, 2) find fulfilling social interactions and a community through streaming to an audience, 3) have an effect on development skills and audience development knowledge, and 4) pose interpersonal and platform related challenges that need to be overcome. We situate our findings in similar streaming and software development research and summarize motivations, streamers' perceptions of audience members, perceptions of skill improvement, and challenges the streamers face. We also outline recommendations for \textcolor{black}{streaming platforms} and address streamers' challenges.

\subsection{Motivations and Benefits of Streaming \textcolor{black}{($RQ_1$ \& $RQ_3$)}}
\vspace{0.1cm}\noindent\textbf{\emph{Motivations --}} We observed that the motivations to stream are personal and social in nature, \textcolor{black}{often focusing on accountability, self-education, mentoring, visibility of work, and finding a community or those with shared interests.} \textcolor{black}{These motivations lead to streamers perceiving beneficial outcomes of} self-growth and learning. \textcolor{black}{We believe that using live streaming can be a useful tool for developers looking to keep a routine, form habits, and find self-discipline. A number of participants noted that they felt a duty to provide for their viewers and audience, and we speculate that the feelings of being held accountable by others lead to streamers finding positive outcomes and effective learning while streaming.}

\textcolor{black}{Streamers} who use \textcolor{black}{live streaming} as an accountability token to practice and learn new development skills found that they could improve their skills over time (\textcolor{black}{the largest beneficial outcome we found in $RQ_3$.}) \textcolor{black}{Prior research shows think-aloud as a form of learning, and verbally walking through a piece of code or through the thought processes of figuring out a bug potentially benefits both streamer and viewers \cite{van1994think, olson2018thinking}. Streamers noted they can walk through an issue and use chat as a ``rubber duck'' to problem solve and receive feedback on their projects, as well as use the audience as a distributed pair-programming or mob programming entity \cite{zuill2016mob}. Using the audience in these ways and thinking aloud through a problem can be faster than searching through traditional forum posts. This may lead to a more creative solution and skill improvements for the streamer and viewers who participate in the problem-solving process. Think aloud and ``rubber ducking'' during streams may also afford the opportunity for streamers and viewers to show off their own development knowledge and be a fulfilling experience if their solutions correct the issue at hand.}

\textcolor{black}{Self-education appears to be a strong motivator for live streaming that provides beneficial impacts for developers. Streamers using live streaming as a tool to prove themselves as capable was effective for many.} We believe that using live streams as a tool for continued education allows developers to be well-rounded and provides a breakup from the monotony of full-time careers. \textcolor{black}{Additionally, streaming provides a real-time feedback loop for developers and viewers alike that is more instantaneous and self-gratifying than relying on forums or trial-and-error programming approaches. We believe this feedback cycle may be strengthened as streamers gain followers, as viewers who might be familiar with the stream might be more eager or willing to speak up and provide feedback.}

Pivotal moments of change in motivation to live stream were gradual as stream viewership increased. As more viewers showed up, it was inevitable that streamers needed to start being more conscious of how they interact with their audience and adopting moderators for chat interactions. A rise in viewership can take out the intimacy of one-on-one discussions that gained followers and jeopardize the sense of community streamers fostered. Though a rise in viewership may pose interpersonal challenges, we believe that once over the initial change that increased viewership strengthens community bonds and provides additional motivation for streamers. 

\vspace{0.1cm}\noindent\textbf{\emph{Benefits of Online Streaming --}}
\textcolor{black}{Streamers perceived beneficial outcomes of live streams revolve around improved development skills through self-education. Outside of self-education motivations, initial and continued motivations of accountability, mentoring, visibility, and community may factor into positive learning outcomes and, ultimately, skill improvements for streamers. Prior research in creative art live streaming shows the benefits of using a stream as a means of critique, and our results appear to echo this sentiment of using viewers for critique and feedback in development streams \cite{fraser2019sharing}. Streamers \textcolor{black}{might have the opportunity to} lean on audience members to teach them, with chat members sometimes having domain-specific knowledge to share with the streamer. Most streamers were okay with and encouraged this style of backseat coding, and prior research has shown that forms of pair- and mob programming are as effective as solo programming \cite{demir2021comparison, bryant2008pair, buchan2018leveraging}.  \textcolor{black}{Our data suggests that developers} use streaming to improve their own abilities and to try new things to teach themselves. For \textcolor{black}{some} streamers, the influence of chat does not directly affect \textcolor{black}{their} programming \textcolor{black}{abilities, results may indicate} that the feedback from chat helps improve features and aspects of their software product and game.} 

One participant (P2) spoke about the accountability of streaming was not only to benefit himself and improve his skills and make progress on projects, but also to the benefit of others who watch him. We speculate that accountability for the streamer, who takes a leadership role for their audience, can produce accountability and learn within their sphere of influence \cite{dubinsky2010effective, li2012leadership}. \textcolor{black}{Providing accountability for others in a streamer's sphere of influence may help viewers commit to themselves, provide self-discipline, or help form habits that influence the development skills of those around the streamer.}

Self-education, wanting to learn outside of a formal class or academic media, might influence whether or not they choose to stream. Prior work has shown that self-directed learning for software developers has been seen as essential for creativity and contributing to OSS and is often practiced by those who are employed in a technology company \cite{lemmetty2020self, xu2009volunteers}. \textcolor{black}{The majority (73\%) of our participants identified themselves as self-taught developers, even though they received formal education in computer science or technology-related field, and found that streaming reinforces streamers' self-education habits and desires.} P2 indicated that streaming is his hobby and that he uses streaming to keep his skills to a higher standard, and P12 wanting to learn a new language to be employable in it. We believe that self-education might also be an influence \textcolor{black}{to stream for some,} as the act of putting themselves out there to be watched while learning something new, immersing themselves, being vulnerable to making mistakes, and walking through the solutions and roadblocks can be challenging, yet rewarding. We believe this to be both a motivation factor for streamers and have \textcolor{black}{directly beneficial skill improvement implications} for viewers wanting to learn by watching someone else, shown to be beneficial in prior research \cite{hodges2015we, mayes2002learning, geertshuis2021learning}.

\subsection{Fulfilling Social Interactions and Community \textcolor{black}{($RQ_2$)}}
For the majority of the participants, community factors into their motivations and reasons to continue to stream. The communities surrounding streams are social, technical, and informational in nature. These aspects \textcolor{black}{appear to} lend themselves to being a core element of those who use \textcolor{black}{live streaming} as a learning platform \textcolor{black}{for software and game development}, and have been found to form similarly within Twitch gaming communities \cite{hamilton2014streaming}. Prior work also shows that developers who video blog (vlog) end up forming small communities where viewers come together in shared experiences over the videos developers share \cite{chattopadhyay2021developers}. Streamers found that once they have \textcolor{black}{viewers and ultimately} a community, they were more motivated and eager to stream and looked forward to socializing with their audience. These interactions \textcolor{black}{appear} to be symbiotic with streamers receiving socialization and help from their audience, and their audience receiving socialization and some aspect of learning a new concept or exposure to something new. Community also played a part in information distribution for OSS, as P1 indicated, and \textcolor{black}{as} prior research demonstrates this same importance of community in OSS \cite{van2010importance, mitchell2015networks}. \textcolor{black}{Development live streams may provide a unique approach to} information distribution to community\textcolor{black}{-developed} applications and a way to update non-technical and technical community \textcolor{black}{members} alike with new features \textcolor{black}{and future development work for an application or release cycle.}

Streamers looked forward to, befriended, mentored, and watched their audience learn new ideas and programming concepts ($RQ_2$)  Additionally, the majority of our participants stated they wanted to be developer advocates. By streaming, they want others to see their problem-solving processes, approach to new and unique challenges, and breaking stereotypes of who developers are and why everyone can learn to be a developer. Prior research shows that actions like vlogging can break down these stereotypes and foster healthy \textcolor{black}{developer} communities \cite{chattopadhyay2021developers}. \textcolor{black}{We observed that development streams and streamers, through altruism, giving back knowledge, and finding fulfillment in sharing knowledge may also work towards breaking down stereotypes and encouraging community participation.} Contrary to views of Twitch having unruly and toxic audience members, seen in gaming streams, we observed that most streamers' audiences are present with genuine and positive intentions \cite{poyane2019toxic}. \textcolor{black}{These genuine and positive intentions could be a product of the stream content, but also due to smaller audiences where streamers and viewers have the chance to interact at a conversational pace. However, from our observations, we note that streaming style may change as streams grow in audience numbers. Additional audience members may take concentration and priority away from development to focus more on viewers' needs and discussions, which could lead to streamers feeling like they do not accomplish what they set out to work on. However, most participants felt okay with this shift of priorities as it fills a social need for them, but have expressed that sometimes it can be frustrating not to complete the task they set out to in a reasonable amount of time.} 

We observed that streamers' perceptions of audience members are overwhelmingly positive, and positive interactions are often an essential part of the stream. Streamer and audience relationships in the context of \emph{S\&GD} are symbiotic social structures where the exchange of information flows and social and communal bonds are built. By forming and promoting positive, healthy communities, we speculate that software development streams form a positive feedback loop for streamers and viewers. This feedback loop promotes maintaining or learning new software development skills and positive attitudes toward development. Our work contributes to a deeper understanding of the relationships in developer communities, social interactions, and habits of software developers.

\subsection{Challenges and Recommendations}
\vspace{0.1cm}\noindent\textbf{\emph{Challenges of Streaming --}}
Our last research question ($RQ_4$) sought to understand \textcolor{black}{live streaming platforms} better and help identify challenges that streamers noted as a hindrance to their ability to stream. We observed that the majority of steamers' challenges are platform-specific challenges and interpersonal challenges to a lesser degree. \textcolor{black}{We believe that} interpersonal challenges were not a common issue for a majority of our participants due to the expectations set by the streamer and a self-regulating and self-moderating community that trended toward positive interactions and rejecting negative behaviors. 

However, for participants that did face harassment issues while streaming, it is not an uncommon occurrence on \textcolor{black}{streaming platforms} or within development circles \textcolor{black}{as seen in prior work} \cite{wohn2019volunteer, imtiaz2019investigating, chen2021towards}. We believe that the continued fostering of a positive community, developer advocacy through the representation of all types of streamers and stream types, and streamer desires to mentor could change this trend within \textcolor{black}{software and game development} communities as a whole \cite{seering2017shaping}. Additionally, the inclusion of moderators has shown to be beneficial in alleviating stream challenges, which several of our participants have implemented by recruiting friends or community volunteers~\cite{fraser2019sharing}.  

\vspace{0.1cm}\noindent\textbf{\emph{Platform Recommendations --}}
Prior work has covered several of the issues our streamers also faced related to visibility and discoverability \cite{chattopadhyay2021developers,chen2021towards}. Navigating the \textcolor{black}{streaming platforms} can be daunting, and setting out to make a presence is not aided by the platforms in many ways. \textcolor{black}{We believe that by} providing better visibility and highlighting different types of streams and streamers outside of tags, developers \textcolor{black}{could} reach a broader audience and have the potential to impact and influence new viewers and continue to break down the stereotypes surrounding development. \textcolor{black}{Platforms could partner with developer advocacy groups for special features on the front page and in recommendations, utilize days like Programmer's Day (September 13th) to feature development streams, or partner with well known developer events like Advent of Code\footnote{\textcolor{black}{Advent of Code is a month-long programming puzzle challenge, typically in December of each year, attempted by student and professional developer. https://adventofcode.com/}} to feature educational, follow along streams.} Additionally, while prior research points to recommending \textcolor{black}{streaming platforms to} provide better services for education, our data from streamers indicates that \textcolor{black}{platforms} should make provided tools extensible to be used by the streamers and provide a neutral toolkit and dynamic API for plug-ins and other stream-related automation tools \cite{chen2021towards}. \textcolor{black}{Extensions, APIs, and toolkits from platforms that enable streamers and viewers to link to documentation, definitions, and other relevant contexts within the chat itself can enhance learning experiences without taking away or distracting a streamer from that they are working on and provide quick real time answers to simple or commonly asked questions. These types of integrations and extenstions for chats could also facilitate community interactions and collaborative projects that engage viewers and streamers to develop and program assets and tools for their own community.} Several streamers and their communities already developed and use their own automation mechanisms, and are tailored to their community and personal preferences, while also providing a way to encourage community input and contributions \cite{chen2017codeon, latoza2014microtask}.

\section{Conclusion}

Online tutorials, education, and streaming software and game development have become more popular due to the low barrier to entry for most people. This paper presented a study of 15 software and game development streamers in the United States and internationally. Through these interviews, we identified motivations to stream surrounded personal accountability, promotion of personal projects, self-education, and finding a sense of community. We presented four research questions that helped frame the experiences and perspectives of streamers and their time streaming - identifying the motivations to stream, perceptions of their audience members, perceptions of skill improvement, and challenges of streaming. Most notable to come out of this study is the strong theme and influence of communities surrounding \textcolor{black}{software and game development} streamers. Our research extends recent research on educational live streaming communities \textcolor{black}{by investigating developer's initial and continued motivations to stream, expanding knowledge on the benefits of live streaming communities through contextualizing motivations to stream with beneficial outcomes of streams, and providing data implicating the benefits of development live streaming as a form of alternative education outside of knowledge sharing.}


 \bibliographystyle{elsarticle-num} 
 \bibliography{bipbopbiblio.bib}





\end{document}